\documentclass[showkeys]{revtex4}

\usepackage{epsfig}

\begin{document}

\title{Transverse Averaging Technique for Depletion Capacitance of
      Nonuniform $PN$-Junctions}

\author{Anatoly A. Barybin$^{*}$, and Edval J. P. Santos$^{**}$}
\affiliation{$^{*}$Electronics Department,\\
Saint-Petersburg State Electrotechnical University,\\
197376, Saint-Petersburg, Russia\\
$^{**}$Laboratory for Devices and Nanostructures,\\
Departamento de Eletr\^{o}nica e Sistemas,\\
Universidade Federal de Pernambuco,\\
C.P. 7800, 50670-000, Recife-PE, Brasil\\
E-mail: edval@ee.ufpe.br}

\begin{abstract}
This article evolves an analytical theory of nonuniform
$PN$-junctions by employing the transverse averaging technique
(TAT) to reduce the three-dimensional semiconductor equations to
the quasi-one-dimensional (quasi-1D) form involving all physical
quantities as averaged over the longitudinally-varying cross
section $S(z)$. Solution of the quasi-1D Poisson's equation shows
that, besides the usual depletion capacitance $C_p$ and $C_n$ due to
the $p$- and $n$-layers, there is an additional capacitance $C_s$
produced by nonuniformity of the cross-section area $S(z)$. The
general expressions derived yield the particular formulas obtained
previously for the abrupt and linearly-graded junctions with uniform
cross-section. The quasi-1D theory of nonuniform structures is
demonstrated by applying the general formulas to the $PN$-junctions
of exponentially-varying cross section $S(z)=S_0\exp(\alpha z)$
as most universal and applicable to any polynomial approximation
$S(z)\simeq S_0(1+\alpha z)^n$.
\end{abstract}

%Uncomment for PACS numbers title message
%\pacs{73.40.Kp, 71.10.Ca, 72.80.Cw}
\maketitle

\section{Introduction}
\label{sec:1}

Analytical methods have the advantage of promoting insight into
device behavior, guiding the interpretation of numerical simulations.
However, the analytical theory of semiconductor devices in
terms of rigorous physical equations is basically developed in complete
form only within the framework of the one-dimensional models~\cite{1}.

Allowance for nonuniformity in the doping impurity distribution and
cross-sectional geometry peculiar to various semiconductor devices requires
applying the two- or~three-dimensional models. Their application can yield
practically useful results only if one resorts to numerical calculations.
But the numerical treatment of the sets of partial differential equations
over the two- or three-dimensional domain often proves to be computationally
too intensive even for relatively simple physical models and device
structures. In most instances, the multidimensional treatment, despite its
comprehen\-siveness, gives too much detailed and often redundant information,
which prevents one from clear physical interpretation of numerical
results. Moreover, the pure numerical approach is inferior to analytical
methods in predictability since it is bound up with the specific
parameters and geometrical structure of a device under calculation.

To deal with this problem, we shall search a mathematical description
of complex semiconductor structures that retains the essential physical
properties peculiar to the multidimensional models but enables their
computational complexity to be reduced. Direction of our search is
determined by the following fairly simple
considerations. Any semiconductor device is connected with external
electric circuits by its terminals so that a driving contribution into
the circuit is caused by the internal electronic processes inside the whole
bulk of a device under examination. Hence, the integral character
of the internal process contributions into the external circuit allows
us to transform the input differential equations in such a way
as to obtain their output form involving some device characteristics
directly related to circuit ones (for example, such as the external
circuit current and the voltage applied to a semiconductor diode).
Such mathematical transformations will be performed by integrating
the three-dimensional equations that govern the internal electronic
processes over the cross section of semiconductor structures and so
be referred to as the {\em transverse averaging technique\/}~(TAT).

Application of the TAT to the fundamental equations of semiconductor
electronics converts them into the one-dimensional form. Such a form
contains the physical scalar quantities (potential, charge density, etc.)
and the longitudinal components of vector quantities (electric field, current
density, etc.) which, being averaged over the cross section $S(z)$, depend
only on the longitudinal coordinate~$z$. Besides, the one-dimensional
equations obtained by using the TAT will also include some contour integrals
along interface lines to take into account the proper boundary conditions
between the different domains of semiconductor structure. Such equations
will be referred to as the {\em quasi-one-dimensional} (quasi-1D) ones.
They are completely equivalent to the initial three-dimensional equations
and in this respect are accurate but they deal solely with the physical
quantities averaged over the cross section of a semiconductor structure.

The transverse averaging approach was first applied in paper~\cite{4a} to
the planar MESFET-structures~\cite{4} in order for the effective boundary
of current channel to be found with taking account of carrier diffusion.
In this paper, the TAT is developed, when applied to the mesa-structures
of nonuniform $PN$-junctions. In addition to the geometrical nonuniformity
given by the cross section $S(z)$, the theory will also allow for the
averaged doping profiles $\bar N_D(z)$ and $\bar N_A(z)$; in so doing the
three functions of $z$ should be assumed as known. Section~2 contains the
general integral relations underlying the transverse averaging technique
and their application to deriving the quasi-1D depletion layer equations
for nonuniform junctions is demonstrated in Section~3. Solution of the
quasi-1D Poisson's equation and general analysis of the depletion-layer
capacitance for the nonuniform $PN$-junctions are set forth in Sections~4
and 5. Section 6 completes the quasi-1D analysis by applying the general
equations to the abrupt and linearly-graded junctions with uniform and
exponentially-varying cross sections.

\setcounter{equation}{0}
\section{Integral Relationships of Transverse Averaging Technique}
\label{sec:2}

In order to apply the transverse averaging technique to the general
three-dimensional equations of semiconductor electronics,
the known integral relations~\cite{5}
\begin{equation}
\int\limits_S (\nabla\cdot{\bf A})\,dS\,=
\frac{d}{dz}\int\limits_S ({\bf e}_z\cdot{\bf A})\,dS\, +
\oint\limits_L ({\bf n}\cdot{\bf A})\,dl,
\label{eq:2.1}
\end{equation}
\begin{equation}
\int\limits_S (\nabla\,\varphi)\,dS =
\frac{d}{dz}\int\limits_S ({\bf e}_z\,\varphi)\,dS +
\oint\limits_L ({\bf n}\,\varphi)\,dl,  \qquad\;\;
\label{eq:2.2}
\end{equation}

%%%%%%%%%%%%%%%%%%%%%%%%%%%%%%%%%%%%%%%%%%%%%%%%%%%%%%%%%%%%%%%%%%%%%%%%%
%%%%%%%%%%%%%%%%%%%%%%%%%%%%%%%%%%%%%%%%%%%%%%%%%%%%%%%%%%%%%%%%%%%%%%%%%
\begin{figure}%[htb]
\epsfxsize=4.5in
\centerline{\epsffile{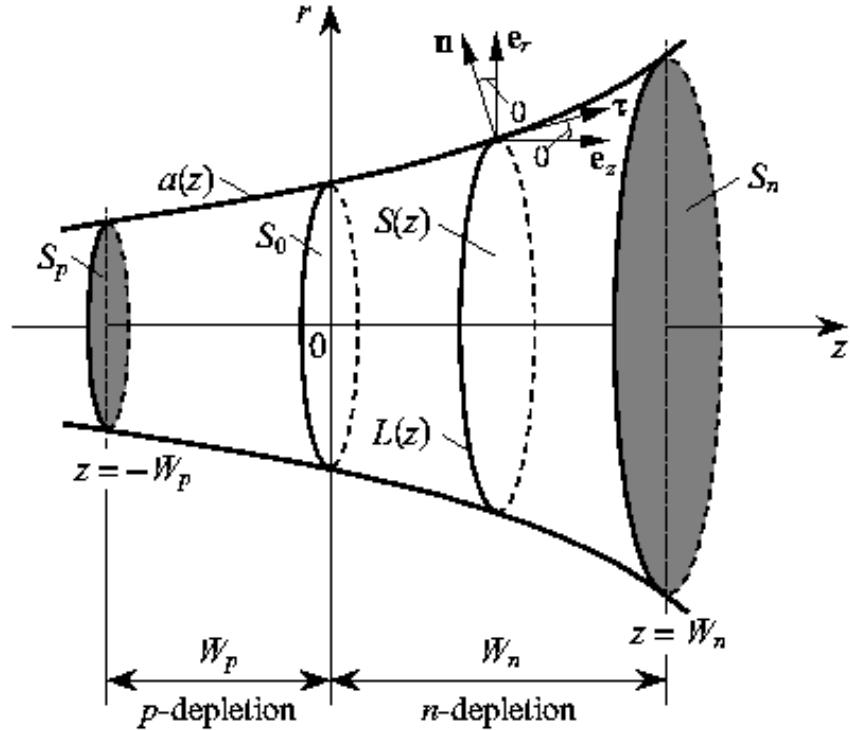}}
\caption{Schematic illustration of the nonuniform $PN$-junction
with axial symmetry; the depletion layers are situated between the
cross sections $z=-W_p$ of area $S_p$ and $z=W_n$ of area $S_n$.}
\label{Fig1}
\end{figure}
%%%%%%%%%%%%%%%%%%%%%%%%%%%%%%%%%%%%%%%%%%%%%%%%%%%%%%%%%%%%%%%%%%%%%%%%%
%%%%%%%%%%%%%%%%%%%%%%%%%%%%%%%%%%%%%%%%%%%%%%%%%%%%%%%%%%%%%%%%%%%%%%%%%

\noindent
written for the uniform cross section $S$ bounded by the contour $L$,
should be generalized to the case of nonuniform structures with
the longitudinally-varying cross section $S(z)$, as shown in Fig.~1.
It is easy to see that the desired generalization of
Eqs.~(\ref{eq:2.1}) and (\ref{eq:2.2}) gives the following forms:
\begin{equation}
\int\limits_{S(z)} (\nabla\cdot{\bf A})\,dS \,=\,
\frac{d}{dz} \,\Bigl[ {\bf e}_z\cdot\bar{\bf A}(z)S(z) \Bigr] +\!
\oint\limits_{L(z)}\!{\frac{{\bf n}\cdot{\bf A}}{\cos\theta}\;dl},
\label{eq:2.3}
\end{equation}
\begin{equation}
\int\limits_{S(z)} (\nabla\,\varphi)\,dS \,=\,
\frac{d}{{dz}} \,\Bigl[ {\bf e}_z\,\bar\varphi(z)S(z) \Bigr] +\!
\oint\limits_{L(z)}\!\frac{{\bf n}\,\varphi}{\cos\theta}\;dl.  \quad\;\;
\label{eq:2.4}
\end{equation}
In Eqs. (\ref{eq:2.3})--(\ref{eq:2.4}), $\bar{\bf A}(z)$ and $\bar\varphi(z)$
are the quantities averaged over the cross-section $S(z)$ (with ${\bf r}$
being the cross-sectional radius-vector,\, see Fig.~1):
\begin{equation}
\bar{\bf A}(z)=
\frac{1}{S(z)}\!\int\limits_{S(z)}\!{\bf A}({\bf r},z)\,dS,
\label{eq:2.6}
\end{equation}
\begin{equation}
\bar\varphi(z)=
\frac{1}{S(z)}\!\int\limits_{S(z)}\!\varphi({\bf r},z)\,dS,
\label{eq:2.7}
\end{equation}
${\bf n}$ is the {\em outward\/} unit vector normal to the boundary
contour $L(z)$, and $\theta$ is the angle of slope of the boundary line
$a(z)$ (characterized by a tangential vector ${\mbox{\boldmath$\tau$}}$)
with respect to the longitudinal unit vector ${\bf e}_z$,\, so that
(see Fig.~1)
\begin{equation}
\cos\theta= \frac{1}{\sqrt{1+(da/dz)^2}}\,.
\label{eq:2.5}
\end{equation}

\setcounter{equation}{0}
\section{Derivation of Quasi-One-Dimensional Depletion-Layer Equations
for Nonuniform $PN$-Junctions}
\label{sec:3}

Electric field inside the depletion layer obeys the usual quasi-static
equations~\cite{1}:
\begin{equation}
\nabla\cdot{\bf D}= \rho \qquad\mbox{with}\qquad
{\bf D}= \epsilon {\bf E},  \quad
\label{eq:3.1}
\end{equation}
\begin{equation}
\nabla\times{\bf E}= 0 \quad\;\;\,\mbox{with}\qquad
{\bf E}= -\nabla\varphi.
\label{eq:3.2}
\end{equation}

Applying the TAT equations~(\ref{eq:2.3}) and (\ref{eq:2.4}) to
Eqs.~~(\ref{eq:3.1}) and (\ref{eq:3.2}), we obtain the following
integral relations for a semiconductor medium:
\begin{equation}
\frac{d}{dz}\,(\bar D_zS) +\!
\oint\limits_{L(z)}\!{\frac{{\bf n}\cdot{\bf D}}{\cos\theta}\;dl} =
\bar\rho\,S,     \qquad
\label{eq:3.3}
\end{equation}
\begin{equation}
\bar E_zS =
-\frac{d}{dz}\,(\bar\varphi S) -\!
\oint\limits_{L(z)}\!
\frac{{\bf n}\cdot{\bf e}_z}{\cos\theta}\;\varphi\, dl.
\label{eq:3.4}
\end{equation}
Here the average longitudinal field and charge density are
defined similar to Eqs.~(\ref{eq:2.6}) and (\ref{eq:2.7}), namely,
\begin{equation}
\bar E_z(z)= \frac{1}{S(z)}\!\int\limits_{S(z)}\!E_z({\bf r},z)\,dS,
\label{eq:3.5}
\end{equation}
\begin{equation}
\bar\rho(z)=
\frac{1}{S(z)}\!\int\limits_{S(z)}\!\rho({\bf r},z)\,dS.  \quad
\label{eq:3.6}
\end{equation}

Outside semiconductor where $\rho\equiv 0$, the equations ana\-logous
to (\ref{eq:3.3}) and (\ref{eq:3.4}) have the following form
(with marking all {\em outside\/} quantities by superscript $o$):
\begin{equation}
\frac{d}{dz}\,(\bar D_z^o S^o) -\!
\oint\limits_{L(z)}\!{\frac{{\bf n}\cdot{\bf D}^o}{\cos\theta}\;dl}= 0,
\qquad\quad
\label{eq:3.7}
\end{equation}
\begin{equation}
\bar E_z^o S^o =
-\frac{d}{dz}\,(\bar\varphi^o S^o) +\!
\oint\limits_{L(z)}\!
\frac{{\bf n}\cdot{\bf e}_z}{\cos\theta}\;\varphi^o dl,
\label{eq:3.8}
\end{equation}
where the outward unit vector for the outside region $S^o$ is
$-{\bf n}$. Here $S^o$ means an effective localization area of the
outside field ${\bf E}^o$ where usually $\epsilon^o\!<\epsilon$. Since
this field is of fringe character, it is reasonable to assume $S^o\!\ll S$.
With taking account of the above inequalities, we can write the
following approximate relations:
\begin{eqnarray}
\bar D_z S + \bar D_z^o S^o &=&
\epsilon\bar E_z S \left( 1+
\frac{\epsilon^o}{\epsilon}\,\frac{\bar E_z^o}{\bar E_z}\,
\frac{S^o}{S} \right) \simeq\,\epsilon\bar E_z S,
\nonumber\\
\bar E_z S + \bar E_z^o S^o &=&
\bar E_z S \left( 1+ \frac{\bar E_z^o}{\bar E_z}\,
\frac{S^o}{S} \right) \simeq\,\bar E_z S,  \qquad\qquad\quad
\label{eq:3.9}\\
\bar\varphi\,S + \bar\varphi^o S^o &=&
\bar\varphi\,S \left( 1+ \frac{\bar\varphi^o}{\bar\varphi}\,
\frac{S^o}{S} \right) \simeq\,\bar\varphi\,S.
\nonumber
\end{eqnarray}

At points of the contour $L(z)$ that demarcates the inside and outside media
the following boundary conditions are valid (with neglecting surface states):
\begin{equation}
{\bf n}\cdot{\bf D}= {\bf n}\cdot{\bf D}^o
\;\;\quad\mbox{and}\quad\;\;   \varphi= \varphi^o.
\label{eq:3.10}
\end{equation}

Because of the boundary conditions (\ref{eq:3.10}), the contour
integrals appearing in Eqs.~(\ref{eq:3.3})--(\ref{eq:3.4}) and
(\ref{eq:3.7})--(\ref{eq:3.8}) cancel each other in pairs after the
respective paired addition of these equations. Then with allowing for
relations (\ref{eq:3.9}), we arrive at the
{\em quasi-one-dimensional equations\/} for the depletion layer of
a nonuniform $PN$-junction in the following form:
\begin{equation}
\frac{d}{dz}\,(\bar E_zS) = \frac{\bar\rho\,S}{\epsilon}\,,
\label{eq:3.11}
\end{equation}
\begin{equation}
\bar E_zS = -\frac{d}{dz}\,(\bar\varphi S).
\label{eq:3.12}
\end{equation}

\setcounter{equation}{0}
\section{Solution of Quasi-One-Dimensional Poisson's Equation}
\label{sec:4}

Substituting (\ref{eq:3.12}) into Eq.~(\ref{eq:3.11}) yields
the {\em quasi-one-dimensional Poisson's equations\/}:

$\bullet$\, for the $n$-depletion layer (where the excess ionized
donors create charge density $\bar\rho=q\bar N_D\equiv
q(\bar N_D^+\!-\bar N_A^-)>0$)
\begin{equation}
\frac{d^2}{dz^2}\,\Bigl[ \bar\varphi(z)\,S(z) \Bigr] =
-\frac{q}{\epsilon}\,\bar N_D(z)\,S(z)
\quad\mbox{for}\quad 0<z<W_n,
\label{eq:3.13}
\end{equation}

$\bullet$\, for the $p$-depletion layer (where the excess ionized
acceptors create charge density $\bar\rho=-q\bar N_A\equiv
-q(\bar N_A^-\!-\bar N_D^+)<0$)
\begin{equation}
\frac{d^2}{dz^2}\,\Bigl[ \bar\varphi(z)\,S(z) \Bigr] =\,
\frac{q}{\epsilon}\,\bar N_A(z)\,S(z)
\,\;\;\quad\mbox{for}\quad\!\! -W_p<z<0,
\label{eq:3.14}
\end{equation}
where the depletion layer widths $W_n$ and $W_p$ (shown in Fig.~1)
should be found.

In order for the quasi-1D Poisson's equations (\ref{eq:3.13}) and
(\ref{eq:3.14}) to be solved, we shall apply the usual boundary
conditions~\cite{1}:

$\bullet$\, for the $n$-depletion layer ($0<z<W_n$)
\begin{equation}
\bar E_z(W_n) = 0 \;\;\qquad\mbox{and}\qquad
\bar\varphi(W_n) = V_n,  \;\;\,
\label{eq:3.15}
\end{equation}

$\bullet$\, for the $p$-depletion layer ($-W_p<z<0$)
\begin{equation}
\bar E_z(-W_p) = 0 \qquad\mbox{and}\qquad
\bar\varphi(-W_p) = V_p,
\label{eq:3.16}
\end{equation}
where $V_n$ and $V_p$ are the voltage drop across the $n$- and
$p$-depletion layers of widths $W_n$ and $W_p$, respectively.

Integration of Eqs.~(\ref{eq:3.13}) and (\ref{eq:3.14}) with allowing
for Eq.~(\ref{eq:3.12}) and the first boundary conditions~(\ref{eq:3.15})
and (\ref{eq:3.16}) imposed on the electric field $\bar E_z$ yields
\begin{equation}
\bar E_z(z)\,S(z) =
-\frac{q}{\epsilon}\,A_n(z) \quad\mbox{for}\qquad\; 0<z<W_n,
\label{eq:3.17}
\end{equation}
\begin{equation}
\bar E_z(z)\,S(z) =
-\frac{q}{\epsilon}\,A_p(z) \quad\mbox{for}\quad -W_p<z<0,
\label{eq:3.18}
\end{equation}
where we have denoted
\begin{equation}
A_n(z)= \int\limits_{z}^{W_n} \!\bar N_D(z')S(z')\,dz',
\label{eq:3.19}
\end{equation}
\begin{equation}
A_p(z)= \!\int\limits_{-W_p}^{z} \!\!\bar N_A(z')S(z')\,dz'.
\label{eq:3.20}
\end{equation}

Integration by parts of Eqs.~(\ref{eq:3.17}) and (\ref{eq:3.18}) with
allowing for Eq.~(\ref{eq:3.12}) and the second boundary conditions
(\ref{eq:3.15}) and (\ref{eq:3.16}) imposed on the potential $\bar\varphi$
yields
\begin{eqnarray}
\bar\varphi(z)S(z) &=& V_n S_n
+ \frac{q}{\epsilon}\,
\bigl[ zA_n(z)- B_n(z) \bigr] \quad\mbox{for}\quad\;\; 0<z<W_n,
\label{eq:3.21}
\end{eqnarray}
\begin{eqnarray}
\bar\varphi(z)S(z) &=& V_p S_p
+ \frac{q}{\epsilon}\,
\bigl[ zA_p(z)- B_p(z) \bigr] \;\quad\mbox{for}\;\; -W_p<z<0,
\label{eq:3.22}
\end{eqnarray}
where we have denoted
\begin{equation}
B_n(z)= \int\limits_{z}^{W_n} \!z'\bar N_D(z')S(z')\,dz', \;\;
\label{eq:3.23}
\end{equation}
\begin{equation}
B_p(z)= \!\int\limits_{-W_p}^{z} \!\!z'\bar N_A(z')S(z')\,dz', \;\;
\label{eq:3.24}
\end{equation}
and (see Fig.~1)
\begin{equation}
S_n\equiv S(W_n),   \qquad   S_p\equiv S(-W_p).
\label{eq:3.25}
\end{equation}

Electric field continuity at $z=0$ imposed on relations (\ref{eq:3.17})
and (\ref{eq:3.18}) in the form
\begin{equation}
\bar E_z(0-)\,=\,\bar E_z(0+)
\label{eq:3.26}
\end{equation}
gives rise to the equality ${A_p(0)=A_n(0)}$, which expresses the
{\em electrical neutrality condition\/} written as
\begin{equation}
\int\limits_{-W_p}^{0} \!\!\bar N_A(z)S(z)\,dz \,=
\int\limits_{0}^{W_n} \!\bar N_D(z)S(z)\,dz.
\label{eq:3.27}
\end{equation}

Potential continuity at $z=0$ imposed on relations (\ref{eq:3.21})
and (\ref{eq:3.22}) in the form
\begin{equation}
\bar\varphi(0-) = \bar\varphi(0+) = 0
\label{eq:3.28}
\end{equation}
gives expressions for the voltage drop across the $n$-
and $p$-layers of depletion written as
\begin{equation}
V_n= \frac{q}{\epsilon S_n}\,B_n(0)=
\frac{q}{\epsilon S_n} \int\limits_{0}^{W_n} \!z\bar N_D(z)S(z)\,dz,
\label{eq:3.29}
\end{equation}
\begin{equation}
V_p= \frac{q}{\epsilon S_p}\,B_p(0)=
\frac{q}{\epsilon S_p} \int\limits_{-W_p}^{0} \!\!z\bar N_A(z)S(z)\,dz.
\label{eq:3.30}
\end{equation}

The total voltage drop across the $PN$-junction consists of the
built-in potential $V_{bi}$ and the applied voltage $V$~\cite{1}:
\begin{equation}
V_n- V_p = V_{bi} - V.
\label{eq:3.31}
\end{equation}
Sign of $V$ is so chosen that the positive ($V\!>0$) and negative ($V\!<0$)
values correspond to the forward-bias and reverse-bias conditions.
From Eqs.~(\ref{eq:3.29})--(\ref{eq:3.31}) it follows that
\begin{equation}
V_{bi}- V =
\frac{q}{\epsilon}\, \biggl[\,
\frac{1}{S_n} \!\int\limits_{0}^{W_n} \!z\bar N_D(z)S(z)\,dz \,-
\frac{1}{S_p} \!\int\limits_{-W_p}^{0} \!\!z\bar N_A(z)S(z)\,dz \,\biggr].
\label{eq:3.32}
\end{equation}

In principle, if the longitudinal dependencies $\bar N_D(z)$, $\bar N_A(z)$,
and $S(z)$ are known, expressions (\ref{eq:3.27}) and (\ref{eq:3.32}) enable
the depletion widths $W_n$ and $W_p$ to be calculated  as functions of
the applied voltage $V$. These expressions constitute a set of the integral
equations where the sought quantities $W_n$ and $W_p$ are the limits of
integration. In some special cases considered below, the above integrals
can be explicitly calculated so that the problem of finding $W_n(V)$ and
$W_p(V)$ takes an algebraic form.

\setcounter{equation}{0}
\section{Depletion-Layer Capacitance of Nonuniform $PN$-Junction}
\label{sec:5}

The expressions (\ref{eq:3.27}) and (\ref{eq:3.32}) allow us to find the
depletion-layer capacitance of the reverse-biased junction (when $V=-|V|<0$)
which is defined as~\cite{1}
\begin{equation}
C(V) = \frac{\partial Q(V)}{\partial |V|}\,.
\label{eq:3.33}
\end{equation}
Here $Q$ is the total charge of excess ionized donors equal to the total
charge of excess ionized acceptors taken in the appropriate layers of
depletion, namely,
\begin{equation}
Q(V)\,=\,
q \!\!\!\!\int\limits_{0}^{W_n(V)}\!\!\!\bar N_D(z)S(z)\,dz
\,=\,q \!\!\!\!\!\!\int\limits_{-W_P(V)}^{0}\!\!\!\!\!\bar N_A(z)S(z)\,dz.
\label{eq:3.34}
\end{equation}

Differentiating expressions (\ref{eq:3.34}), as integrals with variable
limits, with respect to $|V|$, we can redefine the depletion-layer
capacitance (\ref{eq:3.33}) in the following form:
\begin{equation}
C(V) = q\bar N_{Dn}S_n\,\frac{\partial W_n(V)}{\partial |V|}=
q\bar N_{Ap}S_p\,\frac{\partial W_p(V)}{\partial |V|}\,,
\label{eq:3.35}
\end{equation}
where by analogy with notation (\ref{eq:3.25}) we have denoted
\begin{equation}
\bar N_{Dn}\equiv \bar N_{D}(W_n),   \quad\;\;\;
\bar N_{Ap}\equiv \bar N_{A}(-W_p).
\label{eq:3.36}
\end{equation}

Differentiating equation (\ref{eq:3.32}), as integrals with
variable limits $W_n(V)$ and $-W_p(V)$, with respect to $|V|=-V$
(for the reverse-biased junction) and taking into account that,
in accordance with notation~(\ref{eq:3.25}),
\[
S_n(V)= S\bigl( W_n(V) \bigr)
\;\;\quad\mbox{and}\quad\;\;
S_p(V)= S\bigl(-W_p(V)\bigr),
\]
we obtain
\begin{equation} \!\!\!\!\!\!\!\!\!\!\!\!
1= \left( \frac{q}{\epsilon}\,\bar N_{Dn}W_n -
   V_n\,\frac{S'_n}{S_n} \right)\frac{\partial W_n(V)}{\partial |V|} \,+\,
   \left( \frac{q}{\epsilon}\,\bar N_{Ap}W_p -
   V_p\,\frac{S'_p}{S_p} \right)\frac{\partial W_p(V)}{\partial |V|}\,,
\label{eq:3.37}
\end{equation}
where we have denoted
\begin{equation} \!\!\!\!\!\!\!\!\!\!\!\!
S'_n\equiv \frac{dS(z)}{dz} \biggl|_{z\,=\,W_n}\!=
\frac{dS(W_n)}{dW_n}\,,  \qquad
S'_p\equiv \frac{dS(z)}{dz} \biggl|_{z\,=\,-W_p}\!=
-\frac{dS(-W_p)}{dW_p}\,.
\label{eq:3.38}
\end{equation}

Expressing the derivatives $\partial W_n/\partial |V|$ and
$\partial W_p/\partial |V|$ that enter into Eq.~(\ref{eq:3.37}) in
terms of the capacitance $C$ defined by (\ref{eq:3.35}), we finally
arrive at the desired expression for the inverse capacitance $C^{-1}$
of the nonuniform $PN$-junction in the following form:
\begin{equation}
\frac1C= \frac{1}{C_p}+ \frac{1}{C_n}+ \frac{1}{C_s}\,.
\label{eq:3.39}
\end{equation}
In expression (\ref{eq:3.39}), we have introduced the depletion
capacitances for $p$- and $n$-layers:
\begin{equation}
C_p = \frac{\epsilon S_p}{W_p}\equiv\frac{\epsilon S(-W_p)}{W_p}\,,
\qquad
C_n = \frac{\epsilon S_n}{W_n}\equiv\frac{\epsilon S(W_n)}{W_n}\,,
\label{eq:3.40}
\end{equation}
and an additional inverse capacitance $C_s^{-1}$ allowing for nonuniformity
of the cross-section area $S(z)$:
\begin{equation}
\frac{1}{C_s} = -\,\frac{V_p}{q\bar N_{Ap}S_p}\,\frac{S'_p}{S_p} \,-\,
\frac{V_n}{q\bar N_{Dn}S_n}\,\frac{S'_n}{S_n}\,,
\label{eq:3.41}
\end{equation}
where the voltage drops $V_n$ and $V_p$ across the $n$- and $p$-layers
of depletion are given by formulas (\ref{eq:3.29}) and (\ref{eq:3.30}).

From Eq.~(\ref{eq:3.39}) it follows that the resultant depletion
capacitance of a nonuniform $PN$-junction consists of three
capacitances connected in series. Either of the two ones ($C_p$ or $C_n$
given by (\ref{eq:3.40})) presents the capacitance of the planar
capacitor produced by the $p$- or $n$-layer of depletion with the
{\em fixed\/} cross-section area equal to $S_p\equiv S(-W_p)$ or
$S_n\equiv S(W_n)$. The third capacitance $C_s$ defined by
Eq.~(\ref{eq:3.41}) takes into account the cross-section nonuniformity by
the derivatives $S'_n$ and $S'_p$ equal to expressions (\ref{eq:3.38}).

Therefore, for a uniform $PN$-junction with the {\em constant\/}
cross-section area ($S_p(z)=S_n(z)\equiv S_0=$ constant) we have
${C_s^{-1}=0}$. So the differential depletion capacitance
$C=\partial Q/\partial|V|$ is equal to the capacitance $C_0$
for a planar capacitor of the total width $W=W_p+W_n$:
\begin{equation}
C= \frac{C_pC_n}{C_p+C_n} = \frac{\epsilon S_0}{W}\equiv C_0,
\label{eq:3.42}
\end{equation}
which is {\em independent\/} of a character of the longitudinal
distribution for doping impurities $\bar N_A(z)$ and $\bar N_D(z)$.

\setcounter{equation}{0}
\section{Case studies}
\label{sec:6}

Let us analyze the special cases resulting from the general formulas
(\ref{eq:3.27}), (\ref{eq:3.32}), and (\ref{eq:3.39})--(\ref{eq:3.41}), when
applied to the abrupt and linearly-graded junctions. We begin with the~uniform
cross-sectional structures to justify the quasi-1D theory by comparing our
results~with those known from the one-dimensional theory~\cite{1}. The general
quasi-1D theory of~non\-uniform structures is demonstrated by applying
the above formulas to the $PN$-junctions with exponentially-varying cross
section. As will be shown below, the exponential~law $S(z)=S_0\exp(\alpha z)$
is most universal and it can be applied for any polynomial approxi\-mation
$S(z)\simeq S_0(1+\alpha z)^n$.

\subsection{Abrupt junction of uniform cross section:\,
$\bar N_A(z)=$\, constant for $z<0$,\,
$\bar N_D(z)=$\, constant for $z>0$,\,
$S(z)=S_0=$\, constant.}
\label{sec:6.1}

In this case the electrical neutrality condition (\ref{eq:3.27}) and the voltage
relation (\ref{eq:3.32}) respectively yield
\begin{equation}
\bar N_A W_p = \bar N_D W_n,
\label{eq:3.43}
\end{equation}
\begin{equation}
V_{bi}- V= \frac{q}{2\epsilon}\,\bigl( \bar N_D W_n^2+ \bar N_A W_p^2 \bigr).
\label{eq:3.44}
\end{equation}

From Eqs.~(\ref{eq:3.43}) and (\ref{eq:3.44}) it is easy to obtain
the following expressions for the partial width of depletion layers:
\begin{equation}
W_p(V)= \sqrt{\frac{2\epsilon(V_{bi}- V)}{q(\bar N_A+ \bar N_D)}\,
\frac{\bar N_D}{\bar N_A}}\,\equiv
\frac{\bar N_D}{\bar N_A+\bar N_D}\,W(V),
\label{eq:3.45}
\end{equation}
\begin{equation}
W_n(V)= \sqrt{\frac{2\epsilon(V_{bi}- V)}{q(\bar N_A+ \bar N_D)}\,
\frac{\bar N_A}{\bar N_D}}\,\equiv
\frac{\bar N_A}{\bar N_A+\bar N_D}\,W(V),
\label{eq:3.46}
\end{equation}
so that the total depletion width $W=W_p+W_n$ is
\begin{equation}
W(V)= \sqrt{\frac{2\epsilon}{q\bar N}\,(V_{bi}- V)}
\qquad\mbox{with}\quad\;\;
\bar N= \frac{\bar N_A\bar N_D}{\bar N_A+\bar N_D}\,.
\label{eq:3.47}
\end{equation}

Then expression (\ref{eq:3.39}) for the total depletion capacitance
of the uniform $PN$-junction (with $C_s^{-1}\!=0$) takes the form
coincident with formula (\ref{eq:3.42}), namely,
\begin{equation}
C(V)\equiv C_0(V) = \frac{\epsilon\, S_0}{W(V)}=
\sqrt{\frac{\epsilon\, q\bar N}{2(V_{bi}- V)}}\;S_0.
\label{eq:3.48}
\end{equation}

Expressions (\ref{eq:3.47}) and (\ref{eq:3.48}) are fully identical to
those previously obtained for two-sided abrupt junctions~\cite{1}.

For {\em one-sided junctions\/} (e.~g., the $P^+N$-structure with
a highly-doped emitter when $\bar N_A\gg \bar N_D$ so that $W_p\ll W_n$),
expressions (\ref{eq:3.47}) and (\ref{eq:3.48}) assume the following form:
\begin{equation}
W(V)\simeq W_n^0(V) = \sqrt{\frac{2\epsilon}{q\bar N_D}\,(V_{bi}- V)}\,,
\qquad\quad\;\;
\label{eq:3.49}
\end{equation}
\begin{equation}
C(V)\simeq C_n^0(V) = \frac{\epsilon\, S_0}{W_n^0(V)}=
\sqrt{\frac{\epsilon\, q\bar N_D}{2(V_{bi}- V)}}\;S_0,
\label{eq:3.50}
\end{equation}
where superscript 0 characterizes the one-sided uniform junction with
$S_0=$ constant.

\subsection{Linearly-graded junction of uniform cross section:\,
$\bar N_A(z)= -az$ for $z<0$,\,
$\bar N_D(z)= az$ for $z>0$,\,
$S(z)=S_0=$ constant.}
\label{sec:6.2}

In this case the electrical neutrality condition (\ref{eq:3.27}) turns
into an identity because the impurity gradient (equal to $|a|$) is the same
for both the donor and acceptor regions, which ensures symmetry of the
junction with
\begin{equation}
W_p= W_n = \frac W2\,.
\label{eq:3.51a}
\end{equation}

The voltage relation (\ref{eq:3.32}) with allowing for
Eq.~(\ref{eq:3.51a}) takes the following form:
\begin{equation}
V_{bi}- V =
\frac{q}{\epsilon} \!\int\limits_{-W/2}^{W/2} \!\!
z \left[ \bar N_D(z)\!-\!\bar N_A(z) \right]\,dz =\,
\frac{qa}{\epsilon} \!\!\int\limits_{-W/2}^{W/2} \!\!
z^2\,dz =\,\frac{qa}{12\epsilon}\,W^{\,3}.
\label{eq:3.52a}
\end{equation}
From formula (\ref{eq:3.52a}) follows the desired expression for the
total depletion width:
\begin{equation}
W(V)= \left[ \frac{12\epsilon}{qa}\,(V_{bi}-V) \right]^{1/3}\!\!.
\label{eq:3.53a}
\end{equation}

Because of the symmetry relation (\ref{eq:3.51a}), the partial
depletion capacitances for the $p$- and $n$-layers defined by formulas
(\ref{eq:3.40}) are equal to each other:
\begin{equation}
C_p= C_n= \frac{2\epsilon S_0}{W}\,.
\label{eq:3.54a}
\end{equation}
Then formula (\ref{eq:3.42}) gives the total depletion capacitance
of the linearly-graded junction:
\begin{equation}
C(V)= \frac{\epsilon S_0}{W(V)} =
\left[ \frac{qa\epsilon^2}{12(V_{bi}-V)} \right]^{1/3}\!\!S_0.
\label{eq:3.55a}
\end{equation}

Expressions (\ref{eq:3.53a}) and (\ref{eq:3.55a}) are fully identical to
those previously obtained for the linearly-graded junctions~\cite{1}.

\subsection{Abrupt junction of linearly-varying cross section:\,
$\bar N_A(z)=$\,constant for $z<0$,\,
$\bar N_D(z)=$\,constant for $z>0$,\,
$S(z)=S_0(1+ \alpha z)$.}
\label{sec:6.3}

In this case the electrical neutrality condition (\ref{eq:3.27}) takes
the following form:
\begin{equation}
\bar N_A W_p \left( 1- \alpha W_p/2 \right) =
\bar N_D W_n \left( 1+ \alpha W_n/2 \right).
\label{eq:3.56a}
\end{equation}

The voltage drops across the $n$- and $p$-layers of depletion calculated
from Eqs.~(\ref{eq:3.29}) and (\ref{eq:3.30}) by using
$S_n=S_0(1+ \alpha W_n)$ and $S_p=S_0(1- \alpha W_p)$ are equal to
\begin{equation}
V_n = \frac{q\bar N_D W_n^2}{2\epsilon}\:
\frac{1+ (2/3)\alpha W_n}{1+ \alpha W_n}\,,
\label{eq:3.57a}
\end{equation}
\begin{equation}
V_p= -\frac{q\bar N_A W_p^2}{2\epsilon}\:
\frac{1- (2/3)\alpha W_p}{1- \alpha W_p}\,.
\label{eq:3.58a}
\end{equation}

Substitution of Eqs.~(\ref{eq:3.57a}) and (\ref{eq:3.58a}) into equality
(\ref{eq:3.31}) yields an expression that together with Eq.~(\ref{eq:3.56a})
forms the set of algebraic equations to calculate the dependencies $W_p(V)$
and $W_n(V)$ and then to find the depletion capacitance by using
Eqs.~(\ref{eq:3.39})--(\ref{eq:3.41}). However, there is no point in doing
such calculations because the junction of linearly-varying cross section
under consideration is a particular case of the more general situation
corresponding to the junction of exponentially-varying cross section,
which will be proved just below.

\subsection{Abrupt junction of exponentially-varying cross section:\,
$\bar N_A(z)=$\,constant for~$z<0$,\,
$\bar N_D(z)=$\,constant for $z>0$,\,
$S(z)=S_0\exp(\alpha z)$.}
\label{sec:6.4}

In this case the electrical neutrality condition (\ref{eq:3.27}) takes
the following form:
\begin{equation}
\bar N_A \left( 1- e^{-\alpha W_p} \right) =
\bar N_D \left( e^{\alpha W_n}\!- 1 \right).
\label{eq:3.60}
\end{equation}

The voltage drops across the $n$- and $p$-layers of depletion
calculated from Eqs.~(\ref{eq:3.29}) and (\ref{eq:3.30}) by using
$S_n=S_0\exp(\alpha W_n)$ and $S_p=S_0\exp(-\alpha W_p)$ are equal to
\begin{equation} \!\!\!\!\!\!\!\!\!\!\!\!\!\!\!\!\!\!\!\!
V_n= \frac{q\bar N_D W_n^2}{\epsilon\,S_n/S_0}\:
\frac{1- (1\!-\alpha W_n)\exp(\alpha W_n)}{(\alpha W_n)^2} \,=\,
\frac{q\bar N_D}{\epsilon\alpha^2}
\left( e^{-\alpha W_n} +\alpha W_n\!-\! 1 \right),
\label{eq:3.61}
\end{equation}
\begin{equation} \!\!\!\!\!\!\!\!\!\!\!\!\!\!\!\!\!\!\!\!
V_p= -\frac{q\bar N_A W_p^2}{\epsilon\,S_p/S_0}\:
\frac{1- (1+\alpha W_p)\exp(-\alpha W_p)}{(\alpha W_p)^2} \,=
-\frac{q\bar N_A}{\epsilon\alpha^2}
\left( e^{\alpha W_p}\!-\alpha W_p\!-\! 1 \right).
\label{eq:3.62}
\end{equation}

Before applying Eqs.~(\ref{eq:3.60})--(\ref{eq:3.62}) to the subsequent
analysis, let us demonstrate that these equations give rise to the
above formulas (\ref{eq:3.56a})--(\ref{eq:3.58a}) for the abrupt junctions
of linearly-varying cross section. Indeed, when ${\alpha W_p\ll 1}$ and
${\alpha W_n\ll 1}$ so that ${S_n=S_0\exp(\alpha W_n)}\simeq S_0(1+\alpha W_n)$
and $S_p=S_0\exp(-\alpha W_p)\simeq S_0(1\!-\alpha W_p)$,
from Eqs.~(\ref{eq:3.60})--(\ref{eq:3.62}) it follows that
\begin{equation} \!\!\!\!\!\!\!\!\!\!\!\!\!\!\!\!\!\!\!\!
\bar N_A W_p \left( 1- \alpha W_p/2 \right) \,\simeq\,
\bar N_D W_n \left( 1+ \alpha W_n/2 \right),
\label{eq:3.59a}
\end{equation}
\begin{equation} \!\!\!\!\!\!\!\!\!\!\!\!\!\!\!\!\!\!\!\!
V_n= \frac{q\bar N_D W_n^2}{\epsilon\,S_n/S_0}\:
\frac{1-(1\!-\alpha W_n)\exp(\alpha W_n)}{(\alpha W_n)^2} \,\simeq\,
\frac{q\bar N_D W_n^2}{2\epsilon}\:
\frac{1+ (2/3)\alpha W_n}{1+ \alpha W_n}\,,
\label{eq:3.60a}
\end{equation}
\begin{equation} \!\!\!\!\!\!\!\!\!\!\!\!\!\!\!\!\!\!\!\!
V_p= -\frac{q\bar N_A W_p^2}{\epsilon\,S_p/S_0}\:
\frac{1- (1+\alpha W_p)\exp(-\alpha W_p)}{(\alpha W_p)^2} \,\simeq
-\frac{q\bar N_A W_p^2}{2\epsilon}\:
\frac{1- (2/3)\alpha W_p}{1- \alpha W_p}\,.
\label{eq:3.61a}
\end{equation}
Here the following approximate expansions were used:\,
$e^x\simeq 1 + x + x^2/2$\, for Eq.~(\ref{eq:3.59a}) and
$e^x\simeq 1+x+x^2/2+x^3/6$\, for Eqs.~(\ref{eq:3.60a})--(\ref{eq:3.61a}),
where $x=-\alpha W_p$ \,or\, $x=\alpha W_n$.

From a comparison of Eqs.~(\ref{eq:3.56a})--(\ref{eq:3.58a}) and
(\ref{eq:3.59a})--(\ref{eq:3.61a}) follows the conclusion about their
perfect coincidence. Moreover, such a conclusion can be generalized to
the polynomial function ${S(z)=S_0(1+\alpha z)^n,\; n=2,3,\ldots}$ because
the power series of the exponential $e^x$ contains all powers of $x$.
Consequently, we can restrict our attention solely to the junction
with exponentially-varying cross section $S(z)=S_0\exp(\alpha z)$, as the
most universal structure, which allows one to realize any desired polynomial
approximation by making choice of the exponent factor $\alpha$.

Now, let us return to the equations (\ref{eq:3.61}) and (\ref{eq:3.62})
just derived above. Their substitution into equality (\ref{eq:3.31}) yields
an expression that together with Eq.~(\ref{eq:3.60}) forms the set of
transcendental equations to calculate the dependencies $W_p(V)$ and $W_n(V)$.

In order to find the total depletion-layer capacitance, first insert
Eqs.~(\ref{eq:3.61}) and (\ref{eq:3.62}) into expression (\ref{eq:3.41})
for $C_s^{-1}$ with taking account of $S'_p/S_p=S'_n/S_n= \alpha$,
then
\begin{equation}
\frac{1}{C_s}= \frac{1}{C_p} \left(
\frac{\exp(\alpha W_p)-\! 1}{\alpha W_p}\,- 1 \right) +
\frac{1}{C_n} \left(
\frac{1\!- \exp(-\alpha W_n)}{\alpha W_n}\, - 1 \right).
\label{eq:3.63}
\end{equation}
Here the partial depletion capacitances for $p$- and $n$-layers
defined by formulas (\ref{eq:3.40}) are equal to
\begin{equation}
C_p = \frac{\epsilon S_p}{W_p}=
\frac{\epsilon S_0}{W_p}\,e^{-\alpha W_p}
\;\;\quad\mbox{and}\quad\;\;
C_n = \frac{\epsilon S_n}{W_n}=
\frac{\epsilon S_0}{W_n}\,e^{\alpha W_n}.
\label{eq:3.64}
\end{equation}

Insertion of expressions (\ref{eq:3.63}) and  (\ref{eq:3.64}) into
Eq.~(\ref{eq:3.39}) finally gives the desired total inverse depletion
capacitance
\begin{equation}
\frac{1}{C}=
\frac{1}{C_p}\,\frac{\exp(\alpha W_p)- 1}{\alpha W_p}\,+\,
\frac{1}{C_n}\,\frac{1- \exp(-\alpha W_n)}{\alpha W_n}\,.
\label{eq:3.65}
\end{equation}

When $\alpha\to 0$, the above formulas turn into the corresponding
formulas (\ref{eq:3.43})--(\ref{eq:3.48}) obtained for the uniform
two-sided junction.

For {\em one-sided junctions\/} (the $P^+\!N$-diode with
$\bar N_A\gg\bar N_D$, $W_p\ll W_n$, and $V_p\ll V_n$), substitution
of Eq.~(\ref{eq:3.61}) into equality (\ref{eq:3.31}) yields
\begin{equation}
V_{bi}- V\simeq \frac{q\bar N_D}{\epsilon\alpha^2}\,
\Bigl( e^{-\alpha W_n}+ \alpha W_n- 1 \Bigr).
\label{eq:3.66}
\end{equation}

From here follows the transcendental equation for finding $W_n(V)$ written as
\begin{equation}
e^{-\alpha W_n}+ \alpha W_n - 1 = \frac{1}{2}\,\bigl( \alpha W_n^0 \bigr)^2,
\label{eq:3.67}
\end{equation}
where $W_n^0(V)$ is a width of the one-sided uniform ($\alpha= 0$)
junction defined by Eq. (\ref{eq:3.49}).

After calculating $W_n(V)$ from Eq.~(\ref{eq:3.67}), formula
(\ref{eq:3.65}) allows one to find the total depletion capacitance:
\begin{equation}
C(V)\simeq C_n\,\frac{\alpha W_n}{1- \exp(-\alpha W_n)} \,=\,
C_n^0(V)\,\frac{\exp(\alpha W_n)}{W_n/W_n^0 -\alpha W_n^0/2}\,,
\label{eq:3.68}
\end{equation}
where $C_n^0(V)=\epsilon S_0/W_n^0(V)$ is a capacitance of the one-sided
uniform ($\alpha= 0$) junction defined by formula (\ref{eq:3.50})
for the cross section area $S_0=$ constant.

%%%%%%%%%%%%%%%%%%%%%%%%%%%%%%%%%%%%%%%%%%%%%%%%%%%%%%%%%%%%%%%%%%
\begin{figure}[htb]
\epsfxsize=4.5in
\centerline{\epsffile{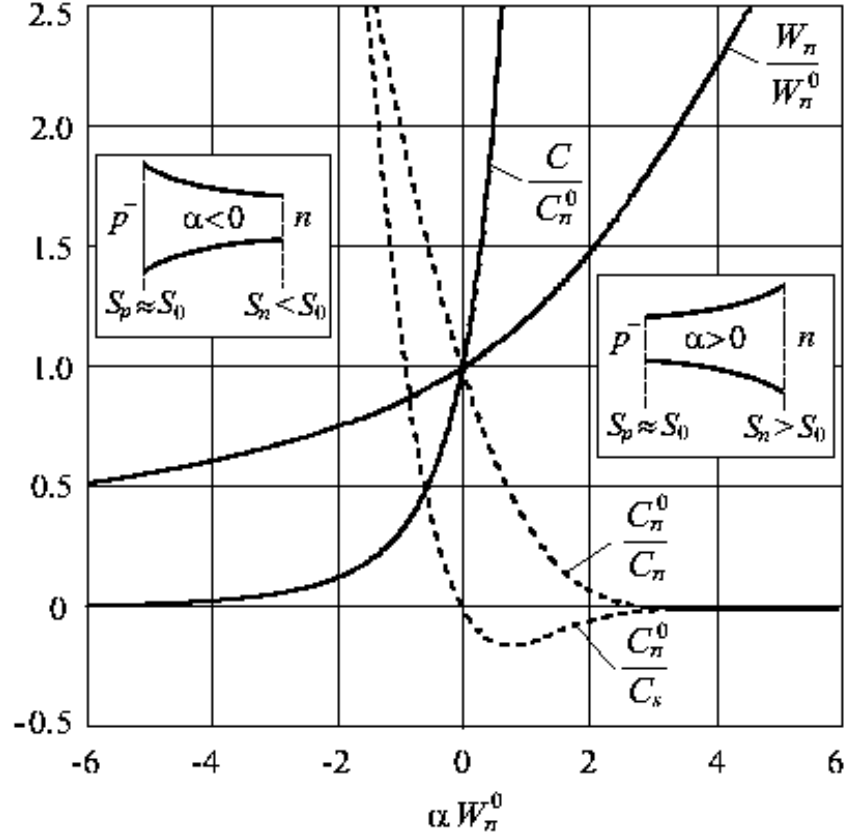}}
\caption{Dependencies of the normalized depletion width $W_n/W_n^0$,
the depletion capacitance $C/C_n^0$ (solid curves), and the normalized
inverse capacitance contributions $C_n^0/C_n$ and $C_n^0/C_s$
(broken curves) versus the nonuniformity parameter~$\alpha W_n^0$
calcu\-lated for the one-sided $P^+\!N$-junction whose nonuniform
geometry are qualitatively shown in the left and right inserts
for $\alpha<0$ and $\alpha>0$.  }
\label{Fig2}
\end{figure}
%%%%%%%%%%%%%%%%%%%%%%%%%%%%%%%%%%%%%%%%%%%%%%%%%%%%%%%%%%%%%%%%

Numerical solution of the transcendental equation (\ref{eq:3.67}) is
carried out for the normalized depletion-layer width $W_n/W_n^0$ as a
function of the nonuniformity parameter $\alpha W_n^0$ and shown in
Fig.~2. Then the normalized depletion capacitance $C/C_n^0$ is calculated
from expression (\ref{eq:3.68}). In Fig.~2, there are also the normalized
inverse capacitance curves $C_n^0/C_s$ and $C_n^0/C_n$ (broken lines)
that give the contributions (\ref{eq:3.63}) and (\ref{eq:3.64}) into
the total capacitance (\ref{eq:3.68}) for the one-sided
$P^+\!N$-junction. They are calculated from the following formulas:
\[
\frac{C_n^0}{C_s} = -\frac{\alpha W_n^0}{2}\,e^{-\alpha W_n}
\quad\,\;\;\mbox{and}\;\;\;\quad
\frac{C_n^0}{C_n} = \frac{W_n}{W_n^0}\,e^{-\alpha W_n}.
\]

As follows from the curves of Fig.~2, the character of cross-sectional
nonuniformity ($\alpha<0$ for $S_p>S_n$ \,or\, $\alpha>0$ for $S_p<S_n$,\,
as shown on the left and right inserts in the figure) produces different
effect on the resultant depletion capacitance. Despite the monotonic
growth of $W_n$ with increasing $\alpha$, the capacitance $C/C_n^0$
grows exponentially for $\alpha>0$ because of increasing an effective
cross-section area when ${S_n>S_p}\simeq S_0$ (see the right insert). For
$\alpha=-|\alpha|<0$,  the capacitance $C/C_n^0$ tends to small values
with growing $|\alpha|$ because of decreasing an effective cross-section
area when $S_n\ll S_p\simeq S_0$ (see the left insert).

\subsection{Linearly-graded junction of exponentially-varying cross section:\,
$\bar N_A(z)=-az$ for~$z<0$,\,
$\bar N_D(z)=az$ for $z>0$,\,
$S(z)=S_0\exp(\alpha z)$.}
\label{sec:6.5}

In this case the electrical neutrality condition (\ref{eq:3.27}) results in
the following expression:
\begin{equation}
(1+\alpha W_p) e^{-\alpha W_p} = (1\!-\alpha W_n) e^{\alpha W_n}.
\label{eq:3.71}
\end{equation}

The voltage drops across the $n$- and $p$-layers of depletion
calculated from Eqs.~(\ref{eq:3.29}) and (\ref{eq:3.30}) by using
$S_n=S_0\exp(\alpha W_n)$ and $S_p=S_0\exp(-\alpha W_p)$ are equal to
\begin{equation}
V_n = \frac{q a}{\epsilon\alpha^3}
\left[ (1\!- 2e^{-\alpha W_n}) + (1\!-\alpha W_n)^2 \right],
\label{eq:3.72}
\end{equation}
\begin{equation}
V_p = \frac{q a}{\epsilon\alpha^3}
\left[ (1\!- 2e^{\alpha W_p}) + (1+\alpha W_p)^2 \right].
\label{eq:3.73}
\end{equation}

Substitution of Eqs.~(\ref{eq:3.72}) and (\ref{eq:3.73}) into
the voltage relation (\ref{eq:3.31}) gives rise to the following:
\begin{equation} \!\!\!\!\!\!\!\!\!\!\!\!
V_{bi}- V = \frac{2q a}{\epsilon\alpha^3} \,\biggl[
\left( e^{\alpha W_p} - e^{-\alpha W_n} \right) +
\alpha^2 (W_n^2-W_p^2)/2 - \alpha (W_n+W_p) \biggr].
\label{eq:3.74}
\end{equation}

Expressions (\ref{eq:3.71}) and (\ref{eq:3.74}) are the transcendental
equations for calculating the dependencies $W_p(V)$ and $W_n(V)$.
It is easy to see that these expressions with $\alpha\to 0$ take the
form of Eqs.~(\ref{eq:3.51a}) and (\ref{eq:3.52a}) obtained before for
the linearly-graded junction of uniform cross section.

The desired expression for the inverse capacitance $C^{-1}$ given by
Eq.~(\ref{eq:3.39}) is obtained by substitution of Eqs.~(\ref{eq:3.72})
and (\ref{eq:3.73}) into formula (\ref{eq:3.41}) for $C_s^{-1}$ with
taking account of\, $S'_p/S_p=S'_n/S_n= \alpha$\, and by the use of
expressions (\ref{eq:3.64}) for $C_p$ and $C_n$, then
\begin{equation}
\frac1C=
\frac{2}{C_p}\:\frac{\exp(\alpha W_p)-\alpha W_p-1}{(\alpha W_p)^2} +
\frac{2}{C_n}\:\frac{\exp(-\alpha W_n) + \alpha W_n - 1}{(\alpha W_n)^2}\,.
\label{eq:3.75}
\end{equation}

When $\alpha\to 0$ and $W_p=W_n=W/2$, formula (\ref{eq:3.75}) assumes
the form of expression (\ref{eq:3.55a}) for the depletion capacitance
of the linearly-graded junction with uniform cross section.

\section{Conclusion}

The paper has demonstrated how to derive the explicit analytic form of the
$C(V)$ characteristics for the $PN$-junctions with nonuniformity in the
cross-sectional geometry and doping impurity distribution by applying the
transverse averaging technique (TAT). Application of the TAT has allowed
semiconductor equations to be reduced to the so-called
{\em quasi-one-dimensional\/} (quasi-1D) form.
Such a form involves all the physical quantities as averaged
over the longitudinally-varying cross section $S(z)$. In~so doing,
the vector quantities (electric field, current density, etc.) retain only
their longitudinal component which, being averaged over $S(z)$, depends
on the longitudinal $z$-coordinate.

The transverse averaging technique has given the quasi-1D Poisson's
equations (\ref{eq:3.13}) and (\ref{eq:3.14}) to analyze the depletion layer
processes. Solution of these equations subject to the boundary conditions
(\ref{eq:3.15}), (\ref{eq:3.16}), (\ref{eq:3.26}) and (\ref{eq:3.28})
has resulted in the integral equations (\ref{eq:3.27}) and (\ref{eq:3.32})
to find the depletion layer widths $W_p(V)$ and $W_n(V)$ as functions of
the applied voltage $V$. The subsequent analysis has shown that the resultant
depletion capacitance $C(V)$ of the nonuniform $PN$-junction consists of
three capacitances connected in series. Either of the two ones ($C_p$ or
$C_n$ given by~(\ref{eq:3.40})) presents the usual capacitance produced by
the $p$- or~$n$-layer of depletion with the fixed cross-section area equal
to $S_p\equiv S(-W_p)$ or $S_n\equiv S(W_n)$. The third additional
capacitance $C_s$ given by~(\ref{eq:3.41}) is due to nonuniformity
of the cross-section area~$S(z)$.

The general expressions has given the known formulas~\cite{1} for the abrupt
and linearly-graded junctions of uniform cross-section, which justifies
the quasi-1D theory developed. When applied to the nonuniform  $PN$-junctions,
the general formulas have been demonstrated for special cases of
the semiconductor mesa-structures with exponentially-varying cross section
$S(z)=S_0\exp(\alpha z)$ as most universal and applicable to any polynomial
approximation $S(z)\simeq S_0(1+\alpha z)^n$. Besides the mesa-structures,
the quasi-1D theory can be also applied to the monolithic planar $PN$-junctions
by considering $S(z)$ as an effective cross section area, which should be found
on the basis of special $C(V)$ measurements for the specific semiconductor
structure under study.

\section{Acknowlegments}

We thank CNPq/``Instituto do Mil\^{e}nio'' Iniative.
One of the authors, AAB, also thanks CNPq for
the support during his stay at UFPE.

\end{document}